\documentstyle[twocolumn,eqsecnum,epsf,aps]{revtex}

\begin{document}
\draft
\title{Inclusive B$\rightarrow$sg decay in QCD}
\author{
I. Halperin$^1$\thanks{E-mail: higor@theory.physics.ubc.ca} and 
M. Lublinsky$^2$\thanks{E-mail:
mal@techunix.technion.ac.il}}

\address{$^1$Department of Physics and Astronomy, University of
 British Columbia, 6224 Agriculture Road, \\ Vancouver, BC V6T 1Z1, Canada}
\address{$^2$Department of Physics, Technion -- Israel Institute of
Technology, Haifa 32000, Israel}
\maketitle

\begin{abstract}
The inclusive nonleptonic transition rate of B meson in gluonic penguin
channel is calculated in QCD using the  
heavy quark mass expansion. We found the branching ratio to be 
approximately  0.25$\%$.
\end{abstract}


\section{Introduction}
\label{sec:intr}

 Recently, there has been a significant interest in the physics  of 
beauty hadrons. From the theoretical viewpoint, its attractiveness is due 
 to the fact that both perturbative and nonperturbative corrections are 
under control
on a scale of the heavy b-quark mass. On the other hand, it is the B-physics 
where the Standard Model can be tested with a high precision and a possible
New Physics is expected to be eventually caught. If a discrepancy between the
predictions of the Standard Model and the data is established, it would be 
necessary to invoke the New Physics for its resolution.

The aim of the present paper is to calculate a contribution to the
 nonleptonic
inclusive decay rate of the B-meson due to the gluon penguin channel 
 (b$\rightarrow$sg).This study is motivated by the existence of the 
semileptonic BR problem \cite{bbsv94}. Despite many theoretical efforts,
this problem so far
remains unsolved within the Standard Model with the 20 \% gap between the data
and theoretical predictions. In attempts to its resolution, it is essential to
 take into
account all relevant contributions which could enhance the BR$_{sl}$.
On the other hand, the gluon penguin contribution to the nonleptonic inclusive
width has not been yet calculated. It is our purpose to fill this gap in 
the literature and calculate the corresponding partial width beyond the 
perturbation theory. We find that no considerable enhancement of BR$_{sl}$ 
occurs due to
this decay  channel, in agreement with the common belief.
Thus, if one is expecting some significant enhancement of the
BR$_{sl}$ in the gluon penguin then an appearance
of a New Physics is anticipated \cite{k95}.

To calculate the inclusive transition rate in the b$\rightarrow$sg channel 
nonperturbatively we
use a powerful method  of pre-asymptotic expansion 
\cite{method,method1,method3}. It is based
upon an incorporation
of elements of the heavy quark effective theory (HQET) \cite{hqet} into
the Wilson
operator product expansion (OPE). In this method, one uses the optical
 theorem to
connect the decay rate $\Gamma$ with the transition operator $ \hat{T} $  
defined by

\begin{equation}
\hat{T}(Q\rightarrow f\rightarrow Q)=i\int{d^4x
T\{H^+_{eff}(x)H_{eff}(0)\}},
\label{e1}
\end{equation}
where $H_{eff}$ denotes an
effective weak Hamiltonian relevant for the $Q\rightarrow f$ transition
. The nonlocal operator $\hat{T}$ is expanded in
a series of local operators with the heavy quark mass serving as a large
parameter of the expansion. The long distance physics is
parametrised by a set of matrix elements of the local operators over the
hadronic states. The transition
rate is calculated in a systematic way through an expansion in inverse
powers of the heavy flavour quark mass.
Thus, inclusive decay rates are determined in terms of this universal set of
matrix elements  without further phenomenological input.

The structure of the paper is as follows. In Sec. 
II we present in short 
the method of pre-asymptotic heavy quark mass expansion. 
The inclusive width in the 
b$\rightarrow$sg channel is calculated in Sec. III.
The final Section IV provides a summary of our results.

\section{pre-asymptotic expansion for inclusive B-decays}

The starting point of our analysis is the expression (\ref{e1}) for the 
transition
operator $\hat{T}$. In the case $b\rightarrow sg$ decay the effective weak
Hamiltonian \cite{gsw88} is
\begin{equation}
H_{eff}(x)=\frac{G_F}{\sqrt{2}}\frac{g}{4\pi ^2}C_W V_{bt} V_{st} s_L (x)
\sigma_{\mu\nu}G_{\mu\nu}(x)b_R(x).
\label{e2}
\end{equation}
Here, $G_F$ is the Fermi constant, $V_{bt}$ and $V_{st}$ stand for the
CKM matrix elements. The effective three point interaction couples the
right-handed b quark with the left-handed s quark and the gluon. The Wilson
coefficient
$C_{W}$ is a result of integrating out hard gluons which generate the
effective vertex (\ref{e2}).
 For the effective Hamiltonian (\ref{e2})
$C_W(m_b)\simeq -0.150$\cite{w96}. The calculation of the decay
rate $\Gamma$ is based on the optical theorem, which relates the inclusive 
width to the
imaginary part of the forward scattering amplitude:
\begin{equation}
\Gamma(B\rightarrow f)=\frac{1}{M_B} Im\langle B|\hat{T} |B\rangle
\label{e3}
\end{equation}
The heavy meson mass $ M_B $ in Eq.(\ref{e3}) accounts for the proper 
normalisation.
 As was pointed out in Ref.
\cite{method,method1,method3},  the expression
(\ref{e3})
suggests a way for calculation of $\Gamma$ through a systematic expansion in
inverse powers of the heavy quark mass. The nonlocal operator $\hat{T}$ is
expanded in a series of local operators, the operators of lowest dimensions
being $O_1=\bar{b}b$ and $O_2=\bar{b}i\sigma Gb$. Thus, $\Gamma$ can be 
written
in the form
\begin{equation}
\Gamma=C_1\langle B|O_1 |B\rangle +C_2 \langle B|O_2|B\rangle + ... .
\label{e4}
\end{equation}
where dots denote contributions of higher order operators. Since these terms
are suppressed by powers of the heavy quark mass their contributions will be
omitted in the following calculations.

In the limit of heavy quark mass, the expectation value of operator $O_1$
over a meson state is
\begin{eqnarray}
\langle B|O_1 |B\rangle=
 \langle B|\bar{b}\gamma_0b|B\rangle+\frac{1}{4m_b^2}
 \langle B|\bar{b}i\sigma Gb|B\rangle  \nonumber \\-
 \frac{1}{2m_b^2}\langle B|\bar{b}\vec{\pi}^2b|B\rangle + O(\frac{1}{m_b^3}).
\label{e5}
\end{eqnarray}
The first term in the expression (\ref{e5}) is equal, up to normalisation,
to the density of b quark in B meson. Taking into account the normalisation
of  the B meson state $\langle B|B\rangle=2M_B$, we get
\begin{equation}
\frac{\langle B|\bar{b}\gamma_0b|B\rangle}{2M_B}=1.
\label{e6}
\end{equation}
The second and third terms in (\ref{e5}) are suppressed by inverse b quark
mass squared. Note, that the second operator is just the $O_2$ operator which
appears in the expansion for the $O_1$ matrix element and plays the role of
chromomagnetic interaction \cite{bs93}. The operator  $O_2$ arises
in the expression
(\ref{e4}) for $\Gamma$ also in a direct way through the expansion of the
transition operator $\hat{T}$. This happens when the interaction of s quark
and gluon with an external gluonic field is taken into account. The relevant
propagators are
\begin{equation}
S(x)=\frac{i\hat{x}}{2\pi^2x^4}+
\frac{x_k\gamma_s\gamma_5}{8\pi^2x^2}\tilde{G}_{ks}(0),
\label{e7}
\end{equation}
\begin{eqnarray}
D_{\alpha\beta}(x-y)&=&\frac{g_{\alpha\beta}}{4\pi^2(x-y)^2}
+\frac{1}{8\pi^2}G_{\alpha\beta}(0)\ln[-(x-y)^2] \nonumber \\
&-&\frac{1}{8\pi^2} G_{\mu\nu}(0) x_\mu y_\nu \frac{g_{\alpha\beta}}{(x-y)^2}.
\label{e8}
\end{eqnarray}
Here $\tilde{G}$ denotes the dual of the field strength.
The matrix element of the chromomagnetic operator is proportional to the
splitting between squared masses of B$^*$ and B mesons:
\begin{equation}
\frac{\langle B|\bar{b}i\sigma Gb|B\rangle}{2M_B}=
\frac{3}{2}(M_{B^*}^2-M_B^2)\simeq 0.7 \  GeV^2
\label{e9}
\end{equation}

The third operator on the right-hand side of equation (\ref{e5}) describes
the kinetic energy of the b quark inside the hadron. Its matrix element
can be estimated \cite{bb94} to be
\begin{equation}
\frac{\langle B|\bar{b}\vec{\pi}^2b|B\rangle}{2M_B}\simeq 0.5\ GeV^2
\label{e10}
\end{equation}

\section{gluon penguin channel in inclusive decay}

\begin{figure}[htbp]
\epsfxsize=0.30\textwidth
\epsfysize=0.30\textwidth
\centerline{\epsffile{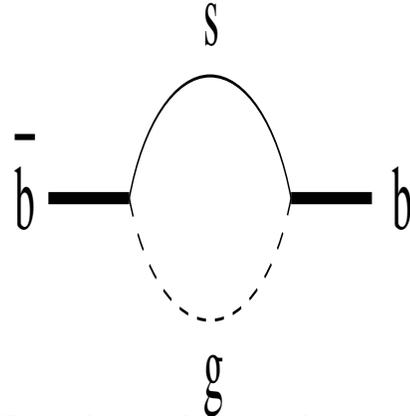}}
\caption{Zero  order contribution to the transition operator $\hat{T}$.}
\label{fig:b1}
\end{figure}

\begin{figure}[htbp]
\epsfxsize=0.44\textwidth
\epsfysize=0.44\textwidth
\centerline{\epsffile{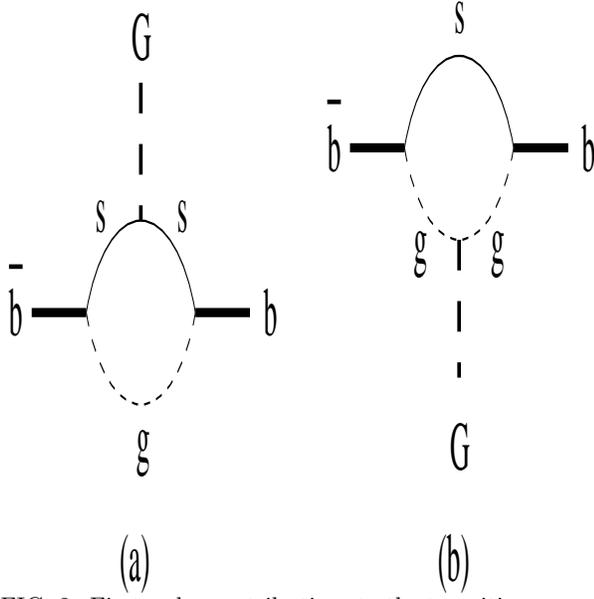}}
\caption{First order contributions to the transition operator $\hat{T}$:
(a)  the contribution of the quark propagator in an external field 
and (b) the contribution of the gluon propagator in an external field.}
\label{fig:b3}
\end{figure}

In order to calculate the inclusive transition rate of B meson in the gluon
 penguin channel, we use
 Eq.(\ref{e1}) with the effective weak Hamiltonian
defined in (\ref{e2}).
\begin{eqnarray}
\hat{T}=Ai\int dx T\{\bar{b}(x)\frac{1-\gamma_5}{2}&&\sigma_{\alpha\beta}
G_{\alpha\beta}(x)s(x) \nonumber \\
\times\bar{s}(0)\sigma_{ij}&&\frac{1+\gamma_5}{2}G_{ij}(0)b(0)\} \nonumber \\
A=\frac{G_F^2}{2}\frac{g^2}{16\pi^4}V_{st}^2V_{bt}^2C_W^2 &&
\label{e11}
\end{eqnarray}
In the limit of the heavy quark mass, the b quark wave function is rescaled
and the mechanical part is extracted \cite{ks83}:
\begin{equation}
\bar{b}(x)=\bar{b}(0)e^{ipx}
\label{e12}
\end{equation}
Here $p$ denotes the four-momentum of the b quark inside the hadron.
The lowest order contributions to the expression (\ref{e11}) for the
transition operator $\hat{T}$ are given by the three diagrams
 (Figs.\ \ref{fig:b1}, \ \ref{fig:b3}).
These diagrams are obtained when the s quark and  gluon fields are
contracted in Eq.(\ref{e11}). Expressions (\ref{e7}), (\ref{e8}) are used for
the propagators in the external gluonic field.

The first diagram  (Fig.\ \ref{fig:b1}) gives rise to the
 operator $O_1$ while the other
two (Fig.\ \ref{fig:b3}) produce the operator $O_2$.
The explicit calculations for the
coefficients  $C_1$ and $C_2$ in (\ref{e4}) yield the following answers:
\begin{eqnarray}
C_1&=&A(\frac{1}{8\pi^2})C_Fm_b^3\ln(-m_b^2) \nonumber \\
C_2&=&A[\frac{1}{(2N_c)8\pi^2}-\frac{N_c}{8\pi^2}]\frac{m_b}{2}\ln(-m_b^2)
\nonumber
\end{eqnarray}
Here, \ $N_c$ stands \  for the number of \ colours and \\
$C_F=(N_c^2-1)/2N_c$. Using the fact that $Im\ln(-m_b^2)=\pi$ we arrive at
the final expression for the inclusive decay rate
\begin{eqnarray}
\Gamma=\frac{G_F^2}{2}\frac{g^2}{16\pi^4}V_{st}^2V_{bt}^2C_W^2\frac{1}{M_B}
[\frac{C_F}{8\pi^2}m_b^3\langle B|\bar{b}b|B\rangle \nonumber \\ +
\frac{m_b}{16\pi^2}(\frac{1}{2N_c}-N_c)\langle B|\bar{b}i\sigma Gb|B\rangle]
\label{e13}
\end{eqnarray}

Substituting expressions (\ref{e5}), (\ref{e6}), (\ref{e9}) and 
(\ref{e10}) for the
matrix elements and using all numerical values we finally determine
the transition rate:
\begin{eqnarray}
\Gamma \simeq 1.1\times 10^{-15} \ GeV.
\nonumber
\end{eqnarray}
The appropriate branching ratio is 2.5$\times10^{-3}$.
\section{Conclusions}
In the present paper, we calculated nonperturbatively the inclusive 
nonleptonic transition
rate of the B meson in the gluonic penguin channel  b $\rightarrow$sg
by means of the pre-asymptotic heavy quark mass
expansion. The obtained BR is about 0.25\% while  the nonperturbative
contribution to the transition rate is about
4\%. Our result provides the Standard Model predictions beyond the
perturbation theory and is to be checked experimentally. The number
obtained means that a significant contribution to the   b $\rightarrow$sg
decay mode is only possible within the New Physics scenarios.

\acknowledgements
We are very  grateful to  Boris Blok  for the suggestion 
of the problem,  useful discussions  during  the  work, and  
reading of the manuscript.

\end{document}